\documentclass[apjl]{emulateapj}

\usepackage{natbib}
\usepackage{longtable}

\slugcomment{To appear in the Astrophysical Journal Letters}

\newcommand{\HII}{H~{\footnotesize II}}

\newcommand{\fluxunits}{erg~s$^{-1}$~cm$^{-2}$}

\newcommand{\ha}{H$\alpha$}
\newcommand{\hbeta}{H$\beta$}

\newcommand{\msun}{$M_\odot$}

\newcommand{\src}{Mrk 709}

\shorttitle{A Massive Black Hole in Mrk 709}
\shortauthors{Reines et al.}

\begin{document}

\title{A Candidate Massive Black Hole in the Low-Metallicity Dwarf Galaxy Pair Mrk~709}

\author{Amy E. Reines\altaffilmark{1}}
\affil{National Radio Astronomy Observatory,
    Charlottesville, VA 22903, USA}
\email{areines@nrao.edu}

\author{Richard M. Plotkin}
\affil{Department of Astronomy, University of Michigan, 500 Church Street, Ann Arbor, MI 48109, USA}

\author{Thomas D. Russell}
\affil{International Centre for Radio Astronomy Research, Curtin University, GPO Box U1987, Perth, WA 6845, Australia}

\author{Mar Mezcua}
\affil{Instituto de Astrof{\'i}sica de Canarias (IAC), E-38200 La Laguna, Tenerife, Spain}
\affil{Universidad de La Laguna, Dept. Astrof{\'i}sica, E-38206 La Laguna, Tenerife, Spain}

\author{James J. Condon}
\affil{National Radio Astronomy Observatory, Charlottesville, VA 22903, USA}
    
\author{Gregory R. Sivakoff}
\affil{Department of Physics, University of Alberta, CCIS 4-181, Edmonton AB T6G 2E1, Canada}

\and  

\author{Kelsey E. Johnson}
\affil{Department of Astronomy, University of Virginia, P.O. Box 400325, Charlottesville, VA 22904-4325, USA}

\altaffiltext{1}{Einstein Fellow}

\begin{abstract}

The incidence and properties of present-day dwarf galaxies hosting massive black holes (BHs) can provide important constraints on the origin of high-redshift BH seeds.  Here we present high-resolution X-ray and radio observations of the low-metallicity, star-forming, dwarf-galaxy system Mrk~709 with the {\it Chandra X-ray Observatory} and the Karl G. Jansky Very Large Array (VLA).  These data reveal spatially coincident hard X-ray and radio point sources with luminosities suggesting the presence of an accreting massive BH ($M_{\rm BH} \sim 10^{5-7}$~\msun).  Based on imaging from the Sloan Digital Sky Survey (SDSS), we find that Mrk~709 consists of a pair of compact dwarf galaxies that appear to be interacting with one another.  The position of the candidate massive BH is consistent with the optical center of the southern galaxy (Mrk~709~S), while no evidence for an active BH is seen in the northern galaxy (Mrk~709~N).  We derive stellar masses of $M_\star \sim 2.5 \times 10^9~M_\odot$ and $M_\star \sim 1.1 \times 10^9~M_\odot$ for Mrk~709~S and Mrk~709~N, respectively, and present an analysis of the SDSS spectrum of the BH-host Mrk~709~S.  At a metallicity of just $\sim$10\% solar, Mrk~709 is among the most metal-poor galaxies with evidence for an active galactic nucleus.  Moreover, this discovery adds to the growing body of evidence that massive BHs can form in dwarf galaxies and that deep, high-resolution X-ray and radio observations are ideally suited to reveal accreting massive BHs hidden at optical wavelengths.

\end{abstract}


\section{Introduction}\label{sec:intro}

In the early universe, the seeds of supermassive black holes (BHs) are believed to have formed in the low-mass, low-metallicity progenitors of today's massive galaxies \citep{silkrees1998,volonteri2010}.  Directly observing the birth and growth of these high-redshift seed BHs is currently not feasible \citep{cowieetal2012,treisteretal2013}, however present-day dwarf galaxies offer another avenue to observationally constrain the origin of supermassive BHs \citep{volonteri2010,greene2012,milleretal2014}. 

\citet{reinesetal2013} recently assembled the largest sample of dwarf galaxies $(M_\star \sim 10^{8.5-9.5}$~\msun) hosting active massive BHs\footnote{Here ``massive BH" refers to BHs more massive than stellar-mass BHs. The median virial BH mass for the broad-line AGN in \citet{reinesetal2013} is $\sim 2 \times 10^{5}$~\msun.} to date using spectroscopy from the Sloan Digital Sky Survey (SDSS).  However, this {\it optically selected} sample of active galactic nuclei (AGN) is biased towards BHs radiating at high fractions of their Eddington luminosities in galaxies with low star formation.  

\begin{figure*}[!t]
\begin{center}$
\begin{array}{cc}
{{\includegraphics[height=3in]{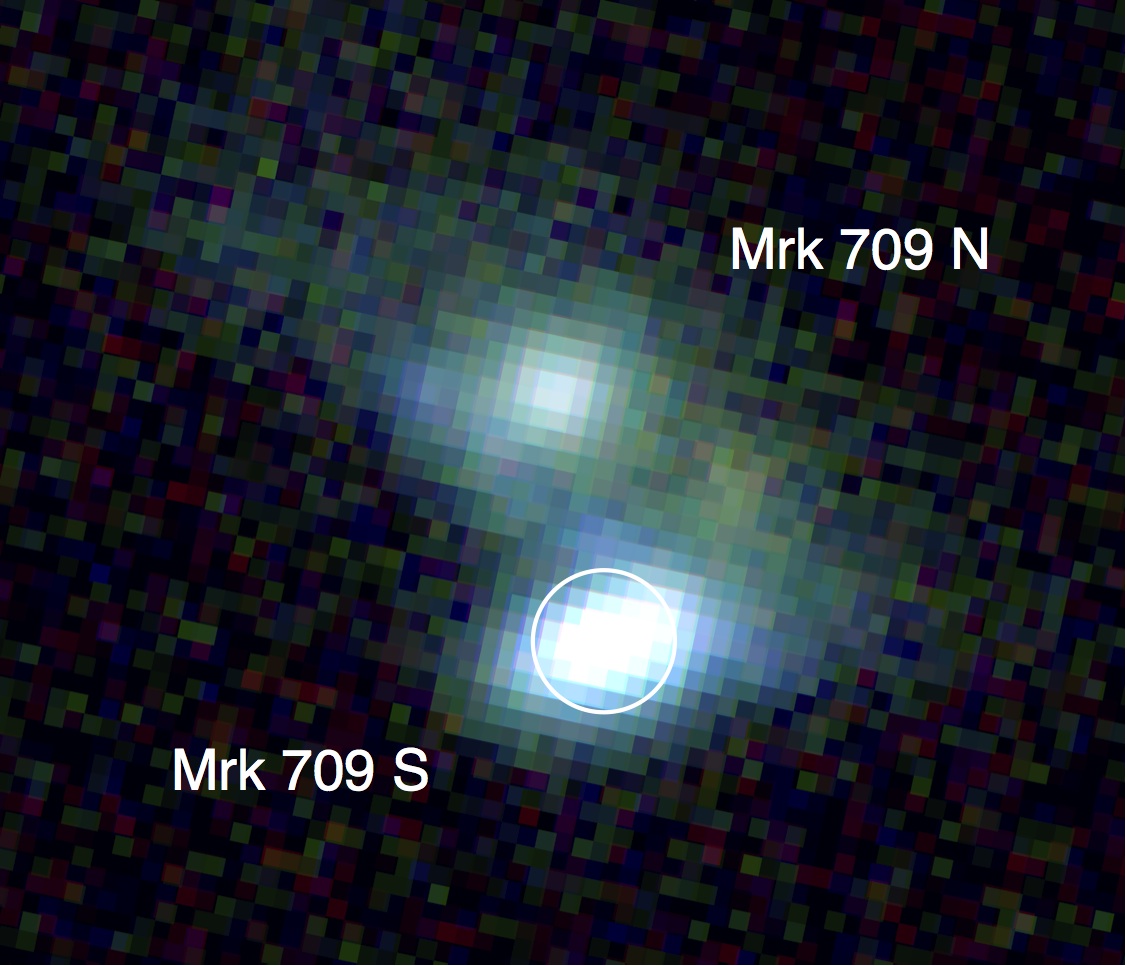}}} &
{{\includegraphics[height=3in]{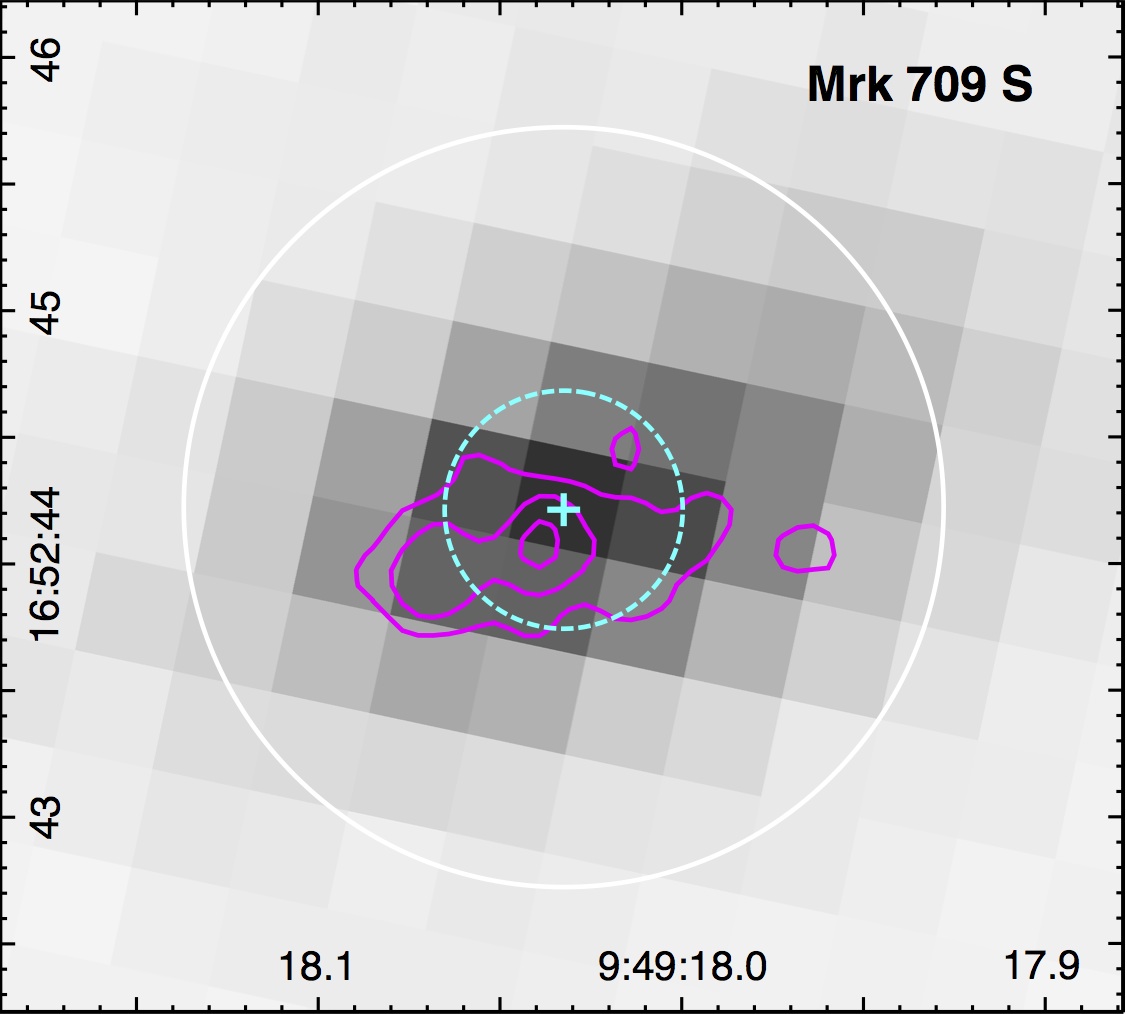}}}
\end{array}$
\end{center}
\caption{\footnotesize {\it Left:}  SDSS image of Mrk~709 (RGB=$zrg$), which appears to be a pair of interacting dwarf galaxies.  We designate the northern and southern galaxies Mrk~709~N and Mrk~709~S.  A logarithmic scaling is used to show extended emission.  The white circle shows the position and size (3\arcsec\ diameter) of the SDSS spectroscopic fiber in both images.  At a distance of $\sim$214 Mpc, 1\arcsec $\sim$1 kpc.  
{\it Right:} SDSS $z$-band image of Mrk~709~S on a linear scale.  The cyan cross and circle mark the position of the {\it Chandra} hard X-ray point source and the 95\% positional uncertainty.  The magenta contours show radio continuum emission at 7.4 GHz from our VLA observations (see Fig.4).  The central radio point source is consistent with the position of the hard X-ray point source.  
\label{fig:sdss}}
\end{figure*}

X-ray and radio observations are more sensitive to BHs with low Eddington ratios \citep{galloetal2008,hickoxetal2009} and can be used to detect massive BHs in actively star-forming dwarf galaxies. % \citep{reinesetal2011,reinesdeller2012}.  
The discovery of the first example of a massive BH in a dwarf starburst galaxy, Henize~2-10, showcases the power of using high-resolution radio and X-ray observations \citep{reinesetal2011,reinesdeller2012}.  
X-ray studies have also uncovered massive BHs in low-mass spheroids \citep{galloetal2008,milleretal2012}, late-type spirals \citep{ghoshetal2008,desrochesho2009}, and low-mass galaxies in the {\it Chandra} Deep Field-South Survey \citep{schrammetal2013}. 

Motivated by our discovery of a massive BH in Henize~2-10 \citep{reinesetal2011}, we have undertaken a combined X-ray and radio search for accreting massive BHs in similar types of galaxies.
%Motivated by our discovery of the first example of a massive BH in a dwarf starburst galaxy, Henize~2-10 \citep{reinesetal2011}, we have undertaken a combined X-ray and radio search for accreting massive BHs in similar types of galaxies.
Here we present {\it Chandra}, VLA, and SDSS observations of Mrk~709, a low-metallicity \citep[$Z \sim 0.1 Z_\odot$, 12 + log(O/H) = 7.7;][]{masegosaetal1994} blue compact dwarf (BCD) galaxy \citep{gildepazetal2003}.  Collectively, these observations strongly suggest the presence of a massive BH in Mrk~709.  

%$ solar is 8.7

Based on SDSS spectroscopy, Mrk~709 has a redshift of $z=0.052$.  We assume $h=0.73$ %$H_0=73$ km s$^{-1}$ Mpc$^{-1}$ 
and adopt of distance\footnote{NED lists the Arecibo H~{\footnotesize I} redshift of z=0.004 from \citet{vanzeeetal1995}.  However, the position listed in that work is $\sim7$\arcmin\ from Mrk~709 and likely refers to another source.} of 214 Mpc to the galaxy.  At this distance, 1\arcsec $\sim1$ kpc.

\section{Observations and Data Reduction}\label{sec:obs}

\subsection{Chandra}\label{sec:xray}

\src\ was observed with \textit{Chandra} for 20.8 ks on 2013 Jan 13 \dataset[ADS/Sa.CXO#obs/13929]{(PI: Reines)}.  The optical center of the galaxy was placed on the S3 chip of the ACIS detector. % \citep{garmireetal2003}.  
We performed data reduction with {\tt CIAO} v4.5. % \citep{fruscioneetal2006}.  
We improved the astrometry by matching X-ray point sources to optical sources ($r<23~$mag) in the SDSS, which has absolute astrometry accurate to $\lesssim$0\farcs1.  The list of X-ray point sources was generated with {\it wavdetect} on a 0.3-7.0 keV image and we found three common X-ray and optical sources, excluding our target.  We registered the \textit{Chandra} image to the SDSS by applying a shift of $\sim$0\farcs06 west and $\sim$0\farcs43 north. Next, we reprocessed the \textit{Chandra} observation (applying caldb v4.5.8).  The entire duration of the observation was considered, as there were no background flares.  
To identify hard X-ray point sources associated with \src, we reran {\it wavdetect} on a 2.0-7.0~keV image of the S3 chip created from the reprocessed event files, using wavelet scales of 1.0, 1.4, 2.0, 2.8, 4.0 pixels, a 5.0~keV exposure map, and a threshold significance of 10$^{-6}$ (approximately one expected false detection on the entire chip). 

\subsection{VLA}\label{sec:vla}

Mrk 709 was observed with the VLA in the most-extended A-configuration on 2012 December 2 (PI: Reines). The observations were taken at C-band with two 1024-MHz basebands centered at 5.0 and 7.4 GHz. Following standard data reduction procedures within CASA,
%the Common Astronomy Software Application \citep[CASA;][]{mcmullinetal2007}, 
3C 286 was used for bandpass and amplitude calibration, while phase calibration was performed using the nearby ($\sim 1.6^{\circ}$ away) compact source J0954$+$1743. Six target observations, each of $\sim$10 minutes, were interspersed between $\sim1.5$ minute phase calibrator observations, and the amplitude and phase gains derived for the calibrator were interpolated to the target source.  

\begin{figure}[!t]
\epsscale{1.2}
\plotone{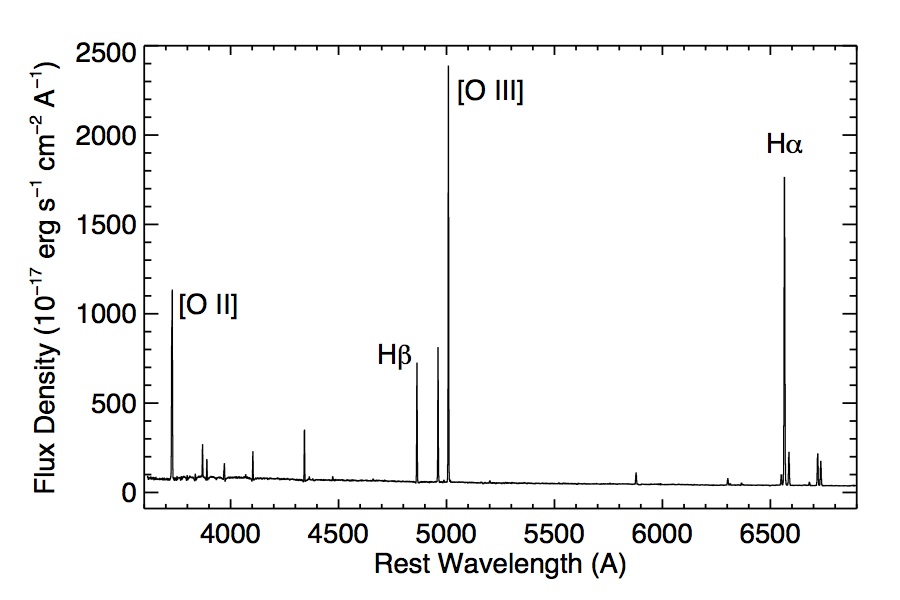}
\caption{\footnotesize SDSS spectrum of Mrk~709 S.}  
\label{fig:spec}
\end{figure}

The calibrated data were then imaged, where deconvolution was carried out with the multi-frequency synthesis algorithm within CASA.  Using natural weighting for maximum sensitivity, we imaged the data at 5.0 and 7.4 GHz.  The 5.0 GHz image has an rms noise of 5.6 $\mu$Jy beam$^{-1}$ and a restoring beam of 0\farcs45 $\times$ 0\farcs41. % (PA = -34.4\degr).  
The 7.4 GHz image has an rms noise of 5.2 $\mu$Jy beam$^{-1}$ and a restoring beam of 0\farcs29 $\times$ 0\farcs28.
%(PA = 82.0\degr).  
At the distance of Mrk~709, the resolution of these observations correspond to linear scales of $\sim$450~pc and $\sim$300~pc at 5.0 and 7.4 GHz, respectively.  The VLA absolute astrometry is accurate to $\lesssim$0\farcs1.

\section{Analysis and Results}

\subsection{Host Galaxy and Star Formation}\label{sec:host}
 
The SDSS image of Mrk~709 (Fig.\ref{fig:sdss}, left) shows a pair of dwarf galaxies likely interacting with one another. The centers of the galaxies are separated by $\sim$5\farcs3, corresponding to a projected physical separation of $\sim$5.5~kpc.  The galaxies appear to share overlapping regions of ionized gas emission that are visible in the SDSS $g$ and $r$ bands due to strong emission lines falling in these bandpasses. % (e.g., \OIII\ and \ha, respectively).  
However, only the southern galaxy, which we designate Mrk~709~S, has an SDSS spectrum (and redshift) so we cannot definitively rule out a chance positional alignment.  The spectroscopic fiber is centered at RA=09$^h$49$^m$18$\fs$031, Dec=+16$^\circ$52$'$44$\farcs$24.

We analyzed the SDSS spectrum of Mrk~709~S to determine if the gas probed by strong emission lines is primarily ionized by massive stars or an AGN (Fig.\ref{fig:spec}).  We used the procedure outlined in \citet{reinesetal2013} to subtract the continuum and absorption lines and measure various emission lines.  A comparison of our measured line ratios with standard diagnostic diagrams \citep{kewleyetal2006} suggests that massive stars are primarily responsible for ionizing the gas.  However, AGN models indicate that line ratios of low-metallicity AGN deviate significantly from Seyfert-like line ratios in high-metallicity systems, and that metal-poor AGN are extremely difficult (if not impossible) to distinguish from low-metallicity star-forming galaxies \citep{grovesetal2006,kewleyetal2013}.  Even if the emission lines in the spectrum of Mrk~709~S are primarily from star formation, this does not rule out the presence of a massive BH.  It simply indicates that star formation is {\it dominating} the narrow-line emission on a galactic scale.  The 3\arcsec\ spectroscopic aperture corresponds to a physical scale of $\sim$3~kpc and covers a significant fraction of this dwarf galaxy (Fig.\ref{fig:sdss}).  

\begin{figure}[!t]
\epsscale{0.9}
\plotone{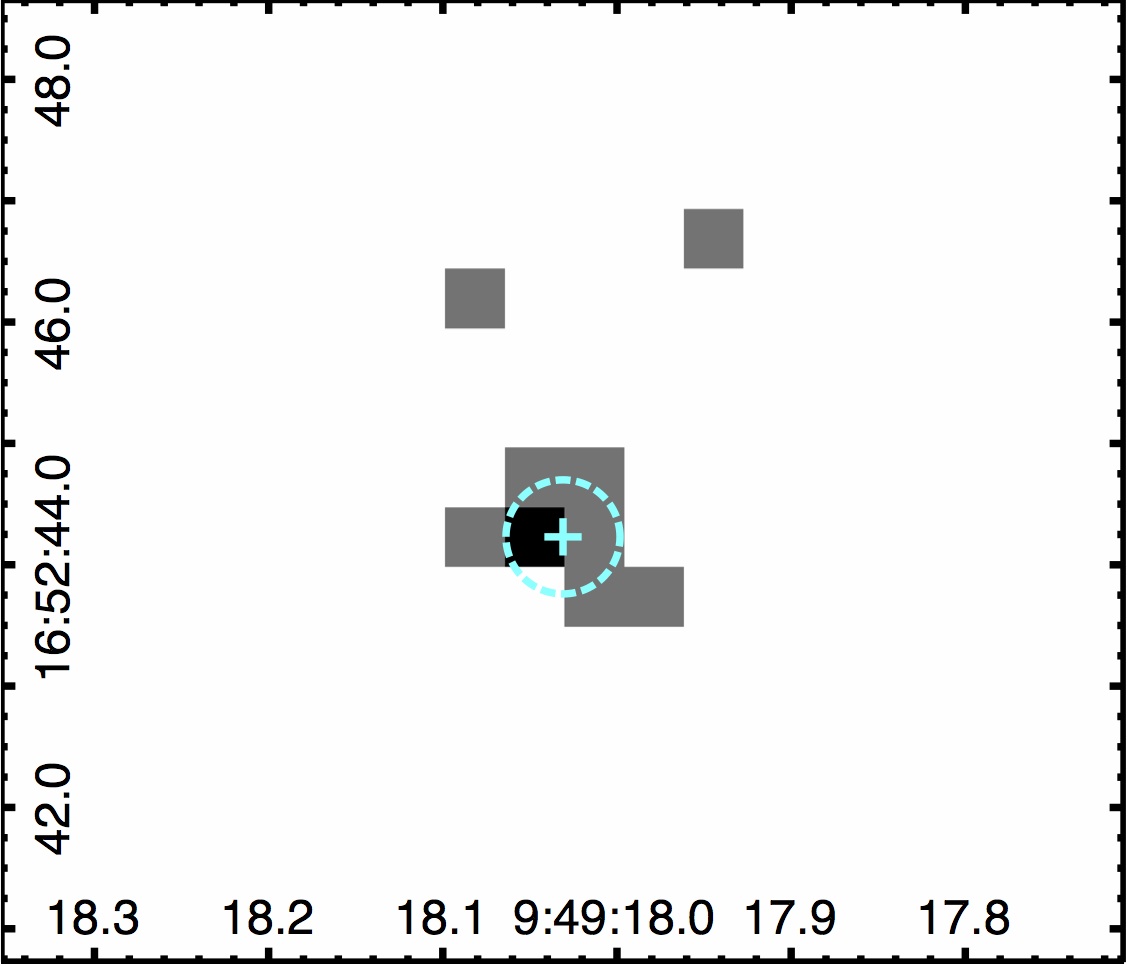}
\caption{\footnotesize {\it Chandra} hard X-ray (2-7 keV) image of Mrk~709~S.  The cross marks the centroid and the circle shows the 95\% positional uncertainty.}
\label{fig:cxo}
\end{figure}

We estimated the star formation rate (SFR) of Mrk~709~S within the 3\arcsec\ spectroscopic aperture using the extinction-corrected \ha\ luminosity and assuming all of the \ha\ emission is due to star formation (and no AGN contribution).  The measured \ha\ flux is $F_{\rm H\alpha} = 8.0 \times 10^{-14}$ \fluxunits\ and the Balmer decrement, \ha/\hbeta\ = 3.47.  This gives an extinction-corrected \ha\ flux of $F_{\rm H\alpha,corr} = 1.3 \times 10^{-13}$ \fluxunits\, corresponding to SFR$_{\rm H\alpha} = 3.9~M_\odot\ {\rm yr}^{-1}$ \citep{kennicuttevans2012}.  The integrated 1.4 GHz flux density from FIRST (3.14 mJy in a 5\arcsec\ beam) corresponds to SFR$_{\rm 1.4GHz} = 10.8~M_\odot\ {\rm yr}^{-1}$.  This is consistent with the \ha-based SFR if multiplied by the ratio of the areas probed, (3\arcsec/5\arcsec)$^2$.

%  Accounting for the larger region probed by FIRST, this is consistent with the \ha-based SFR (i.e., $3^2/5^2 \times 10.8 = 3.9$).

%We estimated the star formation rate (SFR) of Mrk~709~S within the 3\arcsec\ spectroscopic aperture using the extinction-corrected \ha\ luminosity and assuming all of the \ha\ emission is due to star formation (and no AGN contribution).  The measured \ha\ flux is $F_{\rm H\alpha} = 8.0 \times 10^{-14}$ \fluxunits\ and the Balmer decrement, \ha/\hbeta\ = 3.47.  This gives an extinction-corrected \ha\ flux of $F_{\rm H\alpha,core} = 1.3 \times 10^{-13}$ \fluxunits\, corresponding to a luminosity of $L_{\rm H\alpha} = 7.3 \times 10^{41}$ erg s$^{-1}$.  The SFR is then SFR$_{\rm H\alpha} = 10^{{\rm log}(L_{\rm H\alpha}) - 41.27} = 3.9~M_\odot\ {\rm yr}^{-1}$ \citep{kennicuttevans2012}.  The integrated 1.4 GHz flux density from FIRST (3.14 mJy within a region $\sim$ 4\farcs7 $\times$ 4\farcs2) suggests a SFR of 10.8~M_\odot\ {\rm yr}^{-1}$ \citep{kennicuttevans2012}.

%\citep[following][]{dominguezetal2013}

We estimated stellar masses using $z$-band mass-to-light ratios of model galaxies from \citet{gallazzibell2009} and $z$-band fluxes measured using the flexible photometry program SURPHOT \citep{reinesetal2008a}.  Background-subtracted fluxes (in nanomaggies; nMgy) were converted to SDSS magnitudes, and absolute $z$-band luminosities were calculated using $M_{z,\odot} =4.51$ mag \citep{blantonroweis2007}.  From figure 1 of \citet{gallazzibell2009}, model galaxies with young starbursts and low metallicities %(appropriate for Mrk~709) 
have a distribution of $z$-band stellar mass-to-light ratios in the range $-0.2 \lesssim {\rm log}(M_\star/L_z) \lesssim -0.75$, peaking at $\sim -0.55$.  Therefore, $M_\star/L_z \sim 0.3$ with a factor of $\sim$2 uncertainty.  We obtained background-subtracted  
$z$-band fluxes of $284 \pm 6$ and $128 \pm 5$ nMgy for Mrk~709~S and Mrk~709~N, respectively.  The corresponding stellar masses are $M_\star \sim 2.5 \times 10^9$ \msun\ for Mrk~709~S and $M_\star \sim 1.1 \times 10^9$ \msun\ for Mrk~709~N.

We calculated the specific SFR (SFR$/M_\star$) of Mrk 709~S using SFR$_{\rm H\alpha}$ and the $z$-band ``fiberflux" reported in the NASA-Sloan Atlas\footnote{http://nsatlas.org} (169 nMgy).  The stellar mass contained within the 3\arcsec\ spectroscopic aperture, for which we have a corresponding SFR, is $M_\star \sim 1.5 \times 10^9$ \msun.  This gives SFR$/M_\star \sim 2.6 \times 10^{-9}$ yr$^{-1}$.

\subsection{X-ray Emission}\label{sec:xray}

We identified a single hard X-ray {\it Chandra} point source within Mrk~709 (Fig.\ref{fig:cxo}).  The position (RA=09$^h$49$^m$18$\fs$031, Dec=+16$^\circ$52$'$44$\farcs$23) is consistent with the center of Mrk~709~S (Fig.\ref{fig:sdss}).  From equation 5 of \citet{hongetal2005}, we estimated a 95\% positional uncertainty of $0\farcs47$.  

\begin{figure*}[!t]
\begin{center}$
\begin{array}{cc}
{{\includegraphics[width=3.4in]{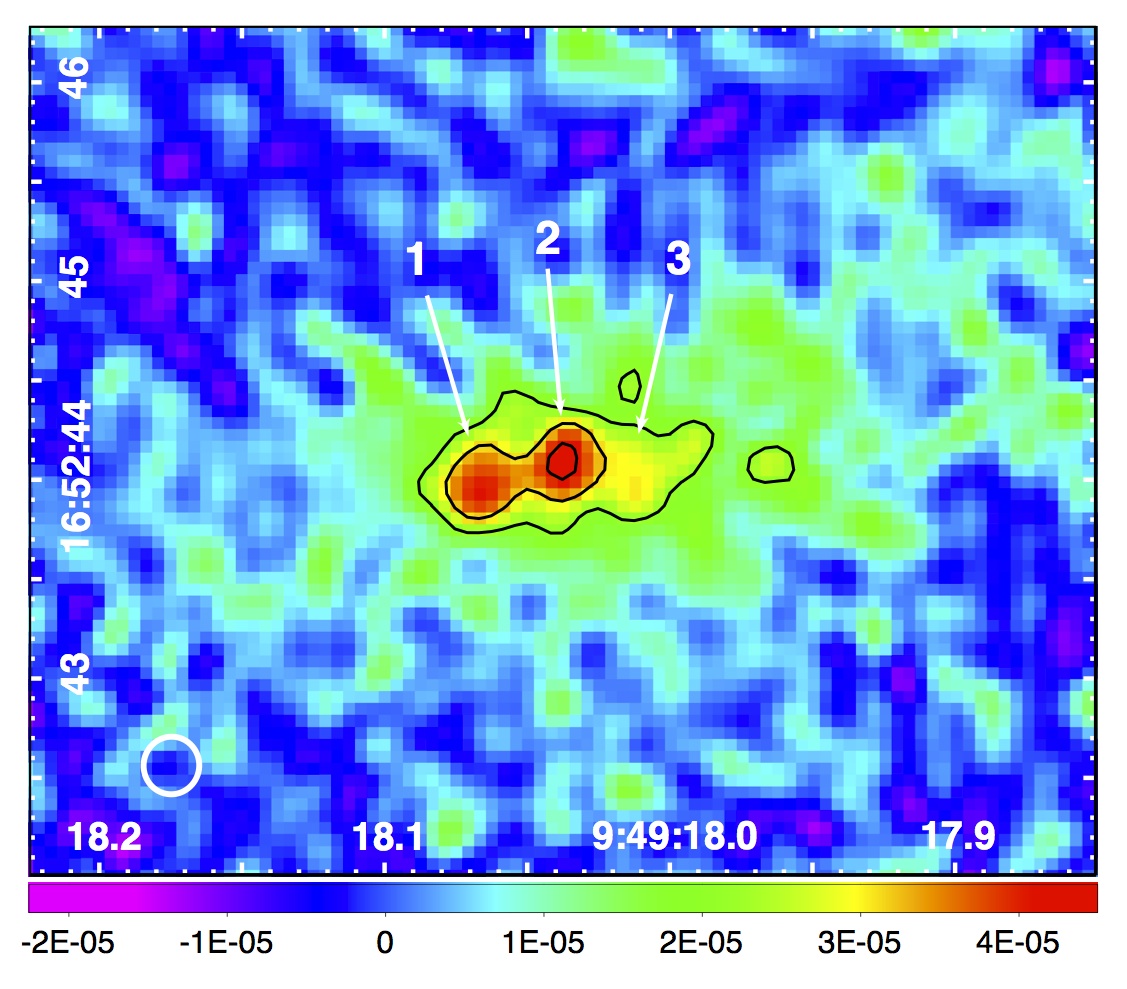}}} &
{{\includegraphics[width=3.4in]{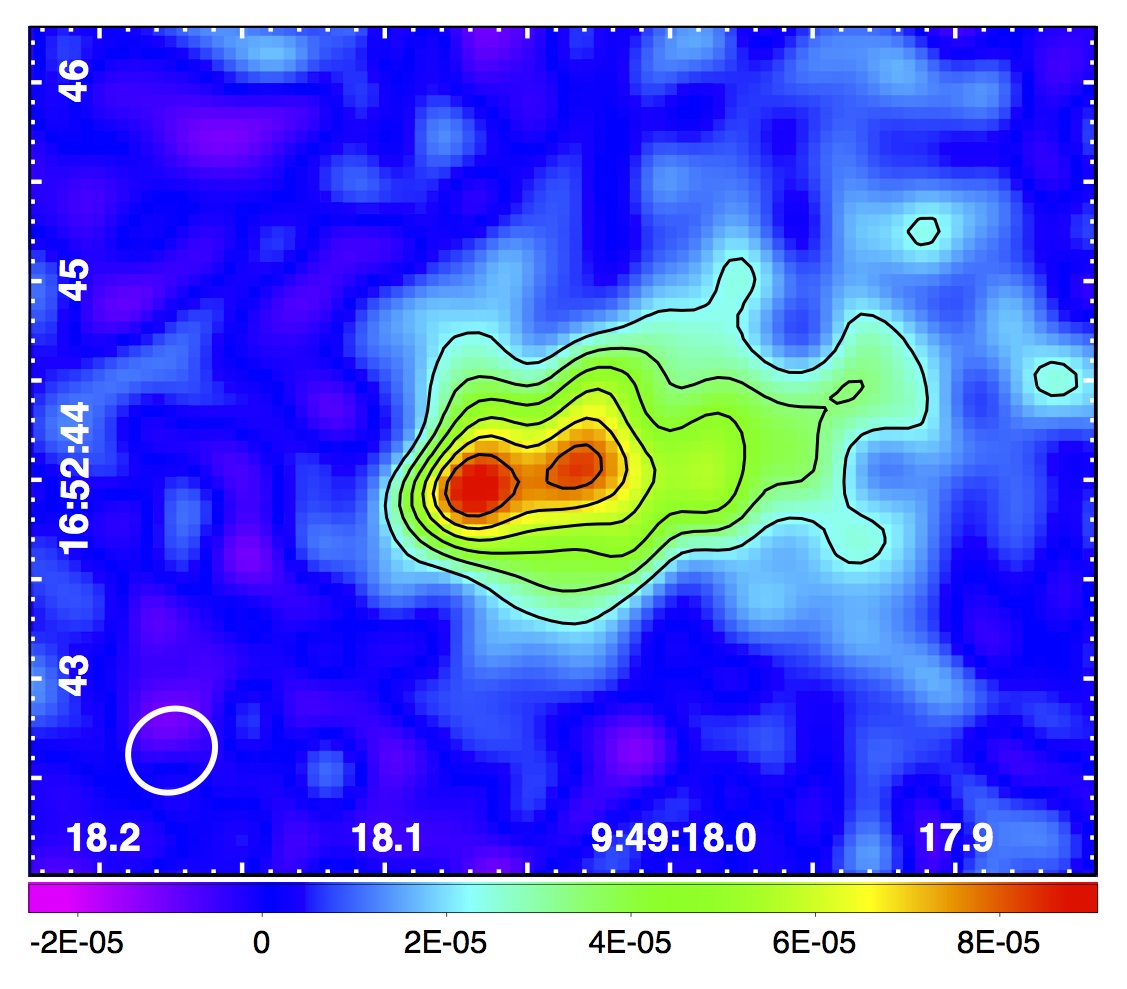}}}
\end{array}$
\end{center}
\caption{\footnotesize VLA 7.4 GHz (left) and 5.0 GHz (right) images of Mrk~709~S. Source 2 has a position consistent with the {\it Chandra} hard X-ray source (Fig.\ref{fig:sdss}).  The scale bars have units of Jy beam$^{-1}$.  Contour levels are 4, 6, 8, 10, 12 and 14 times the rms noise ($\sigma = 5.2$ and 5.6 $\mu$Jy beam$^{-1}$ at 7.4 and 5.0 GHz, respectively).  The beam sizes are shown in the lower left corners. 
\label{fig:vla}}
\end{figure*}

To minimize potential contamination from diffuse X-ray emission from star formation (e.g., hot gas, which primarily emits softer X-rays), we performed X-ray photometry on a hard 2--7~keV \textit{Chandra} image.  We used a circular aperture with a 5$\arcsec$ radius centered on the X-ray source.  Background counts were estimated from a concentric circular annulus with inner and outer radii of 7\arcsec and 12$\arcsec$.  There are 10 hard X-ray counts with an estimated 1.06 background counts within the source's extraction region, constituting a detection at the $>99\%$ confidence level \citep{kraftetal1991}.  Accounting for the background, assuming Poisson statistics and the Bayesian method described in \citealt{kraftetal1991}, we estimated $8.9\pm5.3$ (90\% confidence interval) net X-ray counts from 2-7~keV.   

We obtained the 2-10~keV flux using the net count rates and WebPIMMS, % of the Portable, Interactive Multi-Mission Simulator (PIMMS), 
assuming a power law with photon index $\Gamma=1.8$ and a Galactic absorption $N_{\rm H}=3.14\times10^{20}$~cm$^{-2}$ \citep{dickeylockman1990}.   The corresponding flux is $F_{\rm 2-10 keV} = (9.1 \pm 5.4) \times 10^{-15}$ \fluxunits\ and the luminosity is $L_{\rm 2-10 keV} = (5.0 \pm 2.9) \times 10^{40}$ erg s$^{-1}$.
There could be an additional source of X-ray absorption intrinsic to the source, so these are lower limits.  

\subsection{Radio Emission}\label{sec:radio}
  
Our VLA observations reveal radio emission from Mrk~709~S.  At the resolution of these observations, there is substantial diffuse emission, two relatively bright point-like sources (\#'s 1 and 2 in Fig.\ref{fig:vla}), and a third fainter source (\#3).  
We are primarily interested in the central radio source (\#2), which has a position consistent with the {\it Chandra} hard X-ray point source and the center of Mrk~709~S (Fig.\ref{fig:sdss}).  The offset between source 2 and the hard X-ray source is $\sim$0\farcs17, which is within the astrometric uncertainties.

Radio flux densities were measured using SURPHOT, which allows for a variety of user-defined apertures and backgrounds \citep{reinesetal2008a}.  Given the considerable amount of diffuse emission in Mrk~709~S, the exact choice of aperture and estimate of the background level can lead to large uncertainties.  To account for these effects, we made multiple measurements and incorporated the fluctuations in our errors.  We obtained flux densities of $S_{\rm 7.4 GHz} \sim 40 \pm 10~\mu$Jy and $S_{\rm 5.0 GHz} \sim 60 \pm 20~\mu$Jy for source 2.  Interferometric observations at different frequencies are not sensitive to the same spatial scales and source 2 is not entirely point-like at 5 GHz.  Given these effects, the large errors on the flux densities, and the small spread in frequencies, we are unable to determine if source 2 has a steep or flat radio spectrum between 5.0 and 7.4 GHz.  The radio luminosity of source 2 is $\nu L_\nu = (1.6 \pm 0.6) \times 10^{37}$ erg s$^{-1}$.  

We also imaged the data using different weightings and {\it u-v} cuts to remove diffuse emission and improve the spatial resolution.  Sources 1 and 2 are detected in these images, albeit with signal-to-noise ratios of only S/N $\sim$3.  The peak flux densities in these higher-resolution images are consistent with the flux densities quoted above. 

\subsection{A Candidate Massive Black Hole}\label{sec:bh}

Our {\it Chandra} observations reveal a hard X-ray point source with a luminosity of $L_{\rm 2-10 keV} \geq (5.0 \pm 2.9) \times 10^{40}$ erg s$^{-1}$ at the center of Mrk~709~S.
This luminosity is almost certainly dominated by a single accreting massive BH, as it is significantly higher than the expected contribution from X-ray binaries (XRBs).  For star-forming galaxies with SFR$/M_\star \gtrsim 5.9 \times 10^{-11}$ yr$^{-1}$ (such as Mrk~709~S), high-mass XRBs are expected to dominate the hard (2--10 keV) X-ray emission with the total luminosity of XRBs given by $L_{\rm 2-10 keV} = \alpha M_\star + \beta$SFR, where $\alpha = (9.05 \pm 0.37) \times 10^{28}$ erg s$^{-1}~M_\odot^{-1}$ and $\beta = (1.62 \pm 0.22) \times 10^{39}$ erg s$^{-1}~(M_\odot~{\rm yr}^{-1})$ \citep{lehmeretal2010}.  From Section \ref{sec:host}, $M_\star \sim 1.5 \times 10^9$ \msun\ and SFR $\sim3.9~M_\odot~{\rm yr}^{-1}$ within the spectroscopic fiber. The $3\sigma$ upper limit on the expected contribution from XRBs in this entire 3\arcsec\ region is $L_{\rm 2-10 keV} \sim 9 \times 10^{39}$ erg s$^{-1}$, accounting for the errors in $\alpha$ and $\beta$.  This is a generous upper limit since the expected contribution from XRBs within the {\it Chandra} PSF (1\farcs3 at 3 keV) would be smaller than within the 3\arcsec\ fiber and there could be AGN contamination in our H$\alpha$-based SFR.  Nevertheless, the observed hard X-ray luminosity from the central point source is $\gtrsim$5 times higher than our upper limit on the expected luminosity from XRBs, strongly suggesting the presence of an AGN in Mrk~709~S.  

Assuming Eddington-limited accretion, the {minimum} BH mass is $M_{\rm BH}/M_\odot \ge (\kappa L_{\rm 2-10 keV}) / (1.3 \times 10^{38}$ erg s$^{-1})$, where $\kappa$ is the 2--10 keV bolometric correction.  Even under the most conservative assumptions that all of the luminosity is emitted in the hard X-ray band ($\kappa=1$ and no intrinsic absorption) and the BH is radiating at its Eddington limit, we find $M_{\rm BH} \gtrsim 385$ \msun\ (or $\gtrsim 160$ \msun\ at 95\% confidence given the error on $L_{\rm 2-10 keV}$).  Given that maximally accreting BHs are rare \citep{schulzewisotzki2010} and $\kappa$ is likely $\gtrsim$10 \citep{vasudevanfabian2009}, the BH mass may be orders of magnitude larger.  

%and there may be intrinsic absorption reducing our measured X-ray flux, 

Our VLA observations reveal a point source consistent with the position of the hard X-ray source.  Under the assumption that the radio emission from source 2 is dominated by the X-ray-emitting BH in a low/hard state, we can use the fundamental plane of BH activity \citep{merlonietal2003,falckeetal2004} to obtain an order-of-magnitude estimate of the BH mass and exclude the possibility of a stellar-mass BH at high-confidence.  
We conservatively adopt the $1\sigma$ lower limit on the radio flux density at 5 GHz (40 $\mu$Jy) since source 2 is blended with source 1 at 5 GHz and there may be additional contamination from surrounding diffuse emission.  Using equation 15 in \citet{merlonietal2003} with $F_X = 9.1 \times 10^{-15}$ \fluxunits, $F_R = 2.0 \times 10^{-18}$ \fluxunits, and $D=214$ Mpc, we find $M_{\rm BH} \sim 6 \times 10^6$~\msun.  We caution that this BH mass is highly uncertain due to intrinsic scatter in the fundamental plane (1.1 dex), possible contamination from star formation and/or non-core AGN emission within the VLA beam, and any intrinsic X-ray absorption.  Other fundamental plane relations exist in the literature, which vary due to differing samples and regression techniques \citep[e.g.,][]{falckeetal2004,gultekinetal2009b,plotkinetal2012,millerjonesetal2012}, and they also indicate that if the hard X-ray and radio emission is from an accreting BH, it must be massive (and not a super-Eddington stellar-mass BH).    

\section{Conclusions and Discussion}\label{sec:conclusions}

We have discovered X-ray and radio signatures of an accreting massive BH in the low-metallicity dwarf-galaxy pair Mrk~709.  The position of the BH is consistent with the optical center of the southern galaxy, Mrk~709~S, for which we estimated a stellar mass of $M_\star \sim 2.5 \times 10^9$ \msun\ (comparable to that of the Large Magellanic Cloud).  The central VLA point source (coincident with the hard X-ray {\it Chandra} point source) is straddled by two additional radio sources.  The radio morphology could signify outflow from the central source or circumnuclear star formation.  With a metallicity of $\sim$10\% solar \citep{masegosaetal1994}, Mrk~709 is among the lowest metallicity galaxies with evidence for an AGN \citep{grovesetal2006,izotovthuan2008,ludwigetal2012}.  While local AGNs have primarily been found in massive galaxies with high metallicities \citep[although see][]{reinesetal2013}, low-metallicity (interacting) dwarf galaxies like Mrk~709 were likely more common at high-redshift.   

Alternative explanations for the central co-spatial hard X-ray and radio point sources are difficult to reconcile with our findings.  An ultra-luminous X-ray source powered by a stellar-mass BH could possibly account for the observed X-ray luminosity, $L_{\rm 2-10 keV} \sim 5 \times 10^{40}$ erg s$^{-1}$.  However, a comprehensive {\it Chandra} study of 25 metal-poor galaxies (mostly BCDs) by \citet{prestwichetal2013} found only 6 galaxies had detectable X-ray sources with luminosities $\sim 10-100\times$ lower than the source in Mrk~709~S.  % converted 2-10 keV to .3-8 keV for comparison
Moreover, if the central radio point source emission is also dominated by the BH, the fundamental plane of BH activity firmly rules out a stellar-mass BH.  Alternatives for the radio emission also fall short.  If the central point source were a thermal \HII\ region encompassing young super star clusters, it would require an ionizing luminosity of $Q_{\rm Lyc} \gtrsim 1.7 \times 10^{53}~{\rm s}^{-1}$ \citep{condon1992}, 
 %\citep[from equation 2 in][using our measured flux density at 7.4 GHz and assuming $T=10^4$~K]{condon1992}, 
corresponding to the equivalent of $\gtrsim 17,000$ O7V stars \citep{leitherer1990}.  This would be an extreme \HII\ region, as the ionizing luminosity is significantly larger than that of the thermal radio sources in the metal-poor starburst BCDs SBS 0335-052 \citep{johnsonetal2009} and II Zw 40 \citep{kepleyetal2014}, as well as all of those in the sample of 25 star-forming galaxies by \citet{aversaetal2011}.  The radio luminosity\footnote{We converted our 7.4 GHz flux density to a spectral luminosity at 1.45 GHz assuming a spectral index of $\alpha = -0.7$, where $S \propto \nu^{\alpha}$, for comparison with \citet{chomiukwilcots2009}.  This gives $L_\nu({\rm 1.45~GHz}) \sim 7 \times 10^{27}$ erg s$^{-1}$ Hz$^{-1}$.} would also be exceptionally high for a supernova remnant (SNR), exceeding all radio SNRs in the sample of 18 galaxies by \citet{chomiukwilcots2009}.  In sum, we favor a massive BH origin for the hard X-ray and radio point sources at the center of Mrk~709~S.  

This study underscores the power of utilizing high-resolution X-ray and radio observations to search for accreting massive BHs in low-mass star-forming galaxies that can be missed by optical diagnostics.  Larger-scale surveys with high resolution and sensitivity at radio and X-ray wavelengths are needed to determine how common these objects are, and to ultimately help constrain the BH occupation fraction in dwarf galaxies and the origin of supermassive BH seeds.  Finally, detailed studies of systems like Mrk~709 and Henize 2-10 can teach us about the interplay between BH growth and star formation at low mass, which has implications for high-redshift galaxies \citep{cowieetal2012,treisteretal2013} and contributions to the cosmic X-ray background \citep{xueetal2012}.

\acknowledgements

We thank James Miller-Jones, Jon Miller and Jenny Greene for useful discussions, and the referee for a helpful report.  Support for A.E.R. was provided by NASA through the Einstein Fellowship Program, grant PF1-120086.  This work was also supported by NASA through Chandra Award Number GO2-13126X issued by the Chandra X-ray Observatory Center, which is operated by the Smithsonian Astrophysical Observatory for and on behalf of the NASA under contract NAS8-03060.  The National Radio Astronomy Observatory is a facility of the National Science Foundation operated under cooperative agreement by Associated Universities, Inc.

%SDSS ACKNOWLEGEMENTS.

%\bibitem[{{Greene}(2012)}]{greene2012}
%{Greene}, J.~E. 2012, Nature Communications, 3, 1304

%\bibliography{ref}

\end{document}